\documentclass[twocolumn,secnumarabic,amssymb, nobibnotes, aps, prx,superscriptaddress]{revtex4-2}

\usepackage{etoolbox}
\usepackage{graphicx}  
\usepackage{epstopdf}
\usepackage{dcolumn}   
\usepackage{bm}        
\usepackage{amssymb}   
\usepackage{amsmath}   
\usepackage{amsthm}
\usepackage{chemarrow}
\usepackage{color}
\usepackage{mathrsfs}
\usepackage{float}
\usepackage{bbold}
\usepackage{bbm}

\begin{document}
\title{Implementation of electromagnetic analogy to gravity mediated entanglement}
\date{\today}
\author{Ji Bian}
\affiliation{
	School of Physics and Astronomy, Sun Yat-sen University, Zhuhai 519082, China}
\affiliation{
	Center of Quantum Information Technology, Shenzhen Research Institute of Sun Yat-Sen University, Shenzhen 518087, China}
\author{Teng Liu}
\affiliation{
	School of Physics and Astronomy, Sun Yat-sen University, Zhuhai 519082, China}
\author{Pengfei Lu}
\affiliation{
	School of Physics and Astronomy, Sun Yat-sen University, Zhuhai 519082, China}
\affiliation{
	Center of Quantum Information Technology, Shenzhen Research Institute of Sun Yat-Sen University, Shenzhen 518087, China}
\author{Qifeng Lao}
\affiliation{
	School of Physics and Astronomy, Sun Yat-sen University, Zhuhai 519082, China}
\author{Xinxin Rao}
\affiliation{
	School of Physics and Astronomy, Sun Yat-sen University, Zhuhai 519082, China}
\author{Feng Zhu}
\affiliation{
	School of Physics and Astronomy, Sun Yat-sen University, Zhuhai 519082, China}
\affiliation{
	Center of Quantum Information Technology, Shenzhen Research Institute of Sun Yat-Sen University, Shenzhen 518087, China}
\author{Yang Liu}
\affiliation{
	School of Physics and Astronomy, Sun Yat-sen University, Zhuhai 519082, China}
\affiliation{
	Center of Quantum Information Technology, Shenzhen Research Institute of Sun Yat-Sen University, Shenzhen 518087, China}
\author{Le Luo}
\email[]{luole5@mail.sysu.edu.cn}
\affiliation{
	School of Physics and Astronomy, Sun Yat-sen University, Zhuhai 519082, China}
\affiliation{
	Center of Quantum Information Technology, Shenzhen Research Institute of Sun Yat-Sen University, Shenzhen 518087, China}
\affiliation{
	State Key Laboratory of Optoelectronic Materials and Technologies, Sun Yat-Sen University, Guangzhou 510275, China}
\affiliation{
	International Quantum Academy, Shenzhen, 518048, China}




\begin{abstract}
	Recently, experiments aimed at measuring gravity mediated entanglement (GME) using quantum information techniques have been proposed, based on the assumption that if two  systems get entangled through local interactions with gravitational field, then this field must be quantum. While there is a debate about what could be drawn from GME, quantum simulation might provide some clarification. Here, we present electromagnetic analogy of GME using magnetic-field mediated interaction between the electron and nucleus in a single atom. Our work successfully implements the general procedures of GME experiments and confirms that the mediating field does not support the mean-field description. It also
	clarifies that, without considering the light-crossing time, the GME experiment would not distinguish   a quantum-field-theory description from a quantum-controlled classical field one.
	Furthermore, this work provides  a novel method to construct two-qubit systems in a single atom, and providing  the first quantum simulation of GME using material qubits. It helps to conceive the future  GME experiments on the scale of light-crossing time.
%
\end{abstract}
\maketitle
\section{Introduction}

Understanding gravity within the framework of quantum mechanics is one of the great challenges of modern physics.
Due to the lack of empirical evidence, there is a debate on whether gravity is a quantum entity \cite{oriti2009approaches}. Recently, experiments aimed at measuring quantum gravitational effects, for example, the gravity mediated entanglement (GME) \cite{bose2017spin,marletto2017gravitationally} and the non-Gaussianity \cite{howl2021non}, in table-top experiments using quantum information techniques have been proposed, which might resolve this debate in the end.
The GME proposals are based on a subtle logic that, if two quantum
systems (e.g., two masses that could be prepared in spatial superposition) get entangled through local interactions with a third system (i.e., a gravitational field), then this third system
itself must be quantum \cite{bose2017spin,marletto2017gravitationally}.
Developments in quantum control of larger masses and measurement of gravitational fields of smaller
masses may soon bring this important experimental test into practice \cite{magrini2021real,westphal2021measurement,carney2019tabletop}. At that time, both a positive (the masses are entangled) and a negative result will convey valuable information about quantum gravity \cite{christodoulou2019possibility}.
However, there is also an on-going debate
about the precise conclusions that could be drawn from the detection of GME \cite{christodoulou2019possibility,fragkos2022inference,christodoulou2022locally,martin2022gravity,husain2022dynamics,qiss}. As suggested in Ref. \cite{fragkos2022inference}, GME supports the view that gravitational fields are sourced coherently by  superposition of masses, instead of a mean field description; However, it could not tell whether gravity admits a quantum field description.  Instead, measuring on the light-crossing time between the two masses would truly reveal quantum features of the gravitational interaction, as shown in Ref. \cite{martin2022gravity}.

Based on the detailed correspondence between electromagnetism and
general relativity \cite{maartens1998gravito,cui2021schr}, conducting a GME-like experiment by analogies in electromagnetism could bring additional insights to the above debates. In addition, with the rapid development of quantum information science, quantum simulation of the fundamental properties of spacetime has achieved fruitful results \cite{georgescu2014quantum,jafferis2022traversable,yang2020simulating}. Thus, studies of  the GME experiment using gravito-electromagnetic analogies \cite{maartens1998gravito,costa2014gravito,polino2022photonic} are highly attracting.
In this article, we propose and perform a quantum simulation experiment of the GME effect, using magnetic-field mediated interaction between the electron and nucleus in a single atom.
While the magnetic field is the analog of gravitational field, the spins of electron and nucleus play two roles from the GME experiment. The first role is the position of each mass: different spin states result in different magnetic fields, in analog to the case that different positional combinations of the two masses create different gravitational fields. The second role is the spins possessed by the masses which are used to detect entanglement.
We take quantum field theory (QFT) of electromagnetic field as a premise, and provide a preview of the general experimental procedure of the GME experiment. Observation of entanglement in the final spin state rules out classical mean-field description of the mediating field.
 Moreover, our experimental result confirms that the QFT description and the quantum-controlled classical field one could not be distinguished by original GME proposals, which stresses the necessity of measuring on the order of light-crossing time \cite{martin2022gravity}:
demonstrating that entanglement due to interaction with gravitational field is established between the two masses  at a time smaller than the light-crossing time (entanglement harvesting \cite{pozas2016entanglement}) will help to confirm the quantum nature of the mediating field (with an exception of the yet to be developed gravitational absorber theory \cite{fragkos2022inference}).
We also point out that existing experiments on entanglement between remote trapped-ion qubits \cite{moehring2007entanglement,luo2009protocols,hannegan2022entanglement} could be modified to a simulation of such measurements in electromagnetism.

The experiment utilizes the $d(>2)$-level system (qudit) in a single ion \cite{wang2020qudits,ringbauer2022universal,yuan2022preserving}.
 Moreover, since the four level structure is created by the hyperfine interaction between the nucleus (spin-$1/2$) and the valence electron (spin-$1/2$), the $4$-dimensional Hilbert space is naturally partitioned into two $2$-dimensional subspace. This allows us speak about entanglement of two spins even in a single ion, and makes the experiment easier, as otherwise two trapped ions have to be used to simulate the GME experiment \cite{wang2011quantum,feng2009nuclear}. This work thus demonstrates the advantage that multilevel structure of a quantum unit offers, i.e, simplification of the experimental setup \cite{campbell2022polyqubit,yang2022realizing,gan2020hybrid}, and may shed new light on quantum simulation using trapped ions \cite{monroe2021programmable}.

\section{Realization of the single-atom GME quantum-simulator}\label{realization}

The GME proposal \cite{marletto2017gravitationally,bose2017spin} is illustrated in Fig.\ref{fig01}(a) (see  Appendix \ref{m1}).
 Consider two objects $a$ and $b$ with gravitational interaction, each has mass $m$ and posses spin-$1/2$ , as shown in Fig.\ref{fig01}(a).
Denote the angular momentum operators
\begin{equation*}
I_x= \frac{\hbar}{2}\begin{pmatrix} 0 & 1 \cr 1 & 0 \end{pmatrix}, \
I_y= \frac{\hbar}{2}\begin{pmatrix} 0 & -i \cr i & 0 \end{pmatrix}, \
I_z= \frac{\hbar}{2}\begin{pmatrix} 1 & 0 \cr 0 & -1 \end{pmatrix},
\end{equation*}
and eigenstates of $I_z$ to be $\left|\uparrow \right\rangle$ and $\left|\downarrow \right\rangle$, with corresponding eigenvalues $\hbar/2$ and $-\hbar/2$ (we take $\hbar=1$).
Each mass could be prepared in spatial quantum superposition of two places: up ($u$) and down ($d$), correlated with their spin (e.g., via some Stern-Gerlach scheme). In this paper we presume the interaction is not instantaneous and is mediated by some field.
The derivation of this scenario using a QFT-like framework of gravitational field is already given in, e.g., Ref. \cite{christodoulou2022locally,fragkos2022inference,martin2022gravity}. Here as the light crossing time $t_c \approx d_0/c$ is negligible, a quantum-controlled local classical field derivation would give essentially the same result. In the following we give a Hilbert space to the gravitational field merely to make easy comparison with the electromagnetic simulation.
Denote the state of the whole system as $|\psi_a\rangle \otimes  |\psi_b\rangle \otimes |g\rangle := |\psi_a\rangle   |\psi_b\rangle  |g\rangle$, where $|\psi_{a,b}\rangle$ is the spin state of $a,b$ and $|g\rangle$ represents the gravitational field.
The initial state reads
$
\frac{1}{2}(\left|\uparrow \right\rangle+\left|\downarrow \right\rangle)(\left|\uparrow \right\rangle+\left|\downarrow \right\rangle)|g_{0}\rangle.
$
After the gravitational interaction for time $\tau \gg t_c$, the state becomes
\begin{equation}\label{eq2}
\begin{aligned}
&\frac{1}{2}(e^{-i\phi'_{\uparrow \uparrow}}\left|\uparrow \right\rangle \left|\uparrow \right\rangle |g_{uu}\rangle+e^{-i\phi'_{\uparrow \downarrow}}\left|\uparrow \right\rangle \left|\downarrow \right\rangle  |g_{ud}\rangle \\ &+e^{-i\phi'_{\downarrow \uparrow}}\left|\downarrow \right\rangle \left|\uparrow \right\rangle |g_{du}\rangle+e^{-i\phi'_{\downarrow \downarrow}}\left|\downarrow \right\rangle \left|\downarrow \right\rangle |g_{dd}\rangle),
\end{aligned}
\end{equation}
where
$\phi'_{\uparrow \uparrow}=\Phi_{uu}\tau/\hbar,
\phi'_{\uparrow \downarrow}=\Phi_{ud}\tau/\hbar,
\phi'_{\downarrow \uparrow}=\Phi_{du}\tau/\hbar,
\phi'_{{\downarrow \downarrow}}=\Phi_{dd}\tau/\hbar,$ and $\Phi_{nj}=-Gm^2/d_{nj},$ $(n,j=u,d)$ is the gravitational potential energy when $a$ is in position $n$ and $b$ is in position $j$, $|g_{nj}\rangle$ is the corresponding gravitational field state.
 Finally, one preserves the spin while disentangle the gravitational field, this could be done by coherently bringing $a$ and $b$ back to the middle position in all branches and factor out $|g_0\rangle$. Neglecting the time it takes for this procedure and omitting a global phase, one arrives at the final state (See Appendix \ref{m1})
\begin{equation}\label{gf1}
\frac{1}{2}(e^{-i\phi_{\uparrow \uparrow}}\left|\uparrow \right\rangle \left|\uparrow \right\rangle+\left|\uparrow \right\rangle \left|\downarrow \right\rangle+\left|\downarrow \right\rangle \left|\uparrow \right\rangle+e^{-i\phi_{\uparrow \uparrow}}\left|\downarrow \right\rangle \left|\downarrow \right\rangle)|g_0\rangle,
\end{equation}
where $\phi_{\uparrow \uparrow}=(\Phi_{uu}-\Phi_{ud})\tau/\hbar$.
As long as $e^{-2i\phi_{\uparrow \uparrow}} \neq 1$, $a$ and $b$ are entangled \cite{horodecki2009quantum}.  To detect entanglement, one could utilize Bell inequalities, entanglement witness, or quantum state tomography as explained later. The observation of entanglement will support the view that gravitational fields are sourced coherently by the superposition of sources, and rule out  classical mean field descriptions. In the following, we present  two intuitive analogies of GME in electromagnetism, one involves two electric charges with electric interaction, the other involves two spins with  magnetic interaction. While the first case could be realized by two trapped-ion qubits, the latter is suitable for the single-atom simulation implemented in this work.

 First, we describe the case of two electric charges. Consider the same situation as in Fig.\ref{fig01}(a), and suppose the gravitational interaction and the spin-spin interaction between $a$ and $b$ are too small and could be neglected. But now they both posses electric charges $q$ and interact through electric field. The derivation of this scenario using a QFT description of electromagnetic field is already given in, e.g., Ref. \cite{christodoulou2022locally,fragkos2022inference}. Here as the light crossing time is negligible, a quantum-controlled classical field derivation \cite{martin2022gravity} would give a result with negligible difference. Due to the same reason, a non-relativistic (instantaneous interaction) calculation is further applied merely as a convenient approximation.
 The state of the whole system before final measurement is then
 \begin{equation}\label{e0}
 	\frac{1}{2}(e^{-i\phi_{\uparrow \uparrow}}\left|\uparrow\right\rangle\left|\uparrow\right\rangle+\left|\uparrow\right\rangle\left|\downarrow\right\rangle+\left|\downarrow\right\rangle\left|\uparrow\right\rangle+e^{-i\phi_{\uparrow \uparrow}}\left|\downarrow\right\rangle\left|\downarrow\right\rangle)|E_0\rangle,
 \end{equation}
 where $|E_0\rangle$ represents the electric field when $a$ and $b$ are both in the middle position. $\phi_{\uparrow\uparrow}$ has the same form as \eqref{gf1}, and $\Phi_{nj} (n,j=u,d)$ now represents electric potential energy.
 In this example, the quantum nature (ability to stay in quantum superposition correlated with the source) of electric field is necessary to generate the final entanglement between $a$ and $b$. It is worth mentioning that two-qubit entangling gates in trapped-ion quantum information processors essentially utilize similar mechanism as given above \cite{cirac1995quantum,sorensen1999quantum,kirchmair2009deterministic,leibfried2003experimental,wong2017demonstration,milburn2000ion,blatt2008entangled,bruzewicz2019trapped}. In Sec.\ref{a21}, we provide a reinterpretation of existing experiments \cite{wong2017demonstration} as a simulation of GME.
\begin{figure*}[htb]  
	\makeatletter
	\def\@captype{figure}
	\makeatother
	\centering
	\includegraphics[scale=0.9
	]{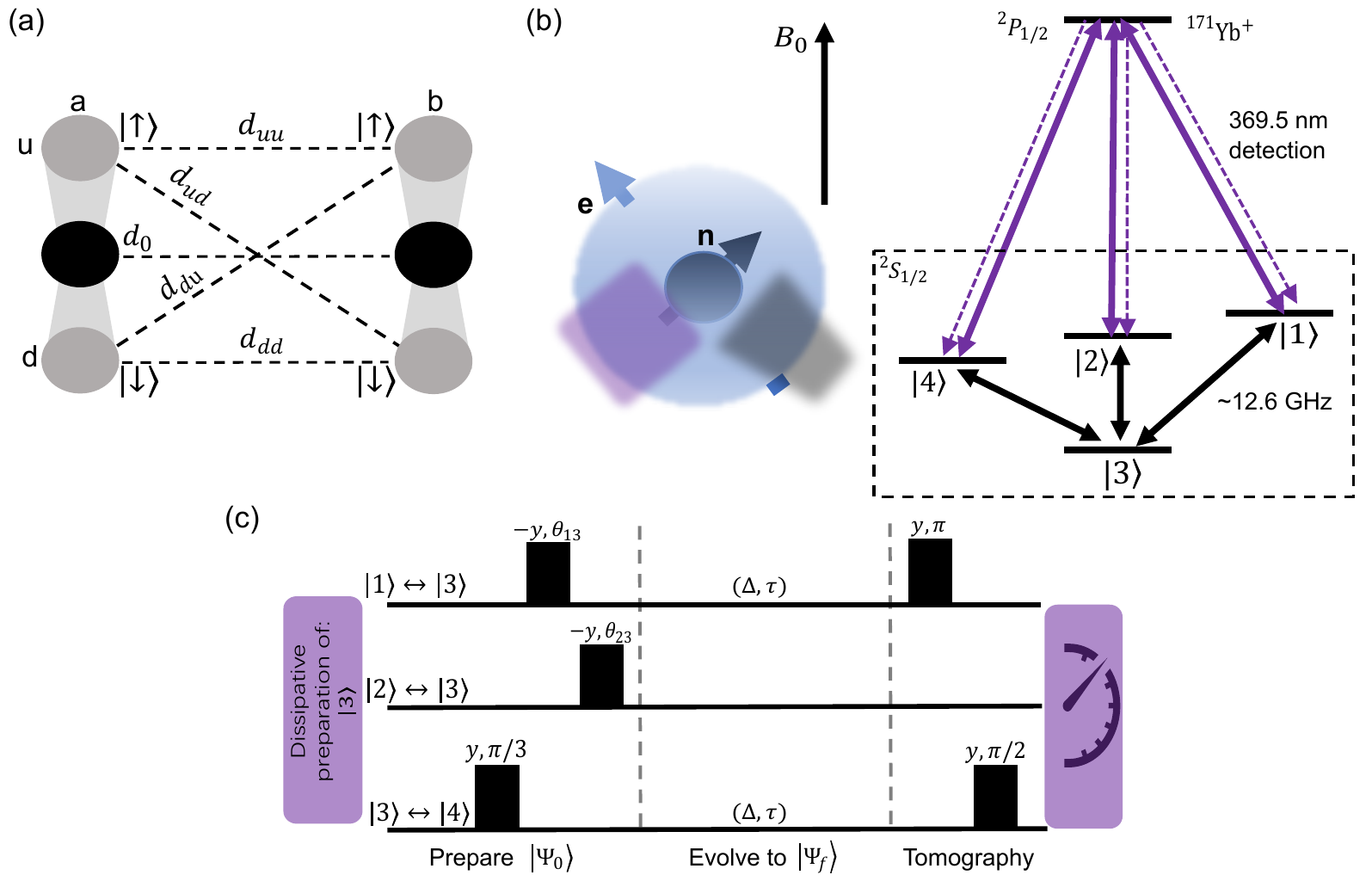}
	\caption{(a) Configuration of the GME experiment considered in this work. The left-right configuration in the original GME proposal \cite{bose2017spin} is modified to be up-down here to simplify the simulation experiment, they are essentially the same. Objects $a$ and $b$ could be put into spatial superposition of $u$ and $d$ correlated with their spin states, and there is gravitational interaction between them. (b) The s-electron (e) in the outermost shell and the nucleus (n) in a Hydrogen-like atom, or in an alkaline-like ion. The electron spin and the nuclear spin interacts through magnetic field and results in the interaction Hamiltonian $H_f$. In the $^{171}$Yb$^{+}$ ion used here, there is a magnetic field $B_0$ in $z$ providing the quantization axis. The corresponding four-level subspace is controlled by microwave fields. The total population on $|1\rangle$, $|2\rangle$, and $|4\rangle$ is measured by applying the $369.5$ nm detection laser beam and measuring fluorescence.
	(c) The experimental pulse sequence in $f_O$ (see Sec.\ref{aexp}). After initializing to $|3\rangle$ by optical pumping, we prepare $|\Psi_0\rangle$ by sequentially applying microwave pulses. We then evolve the system by \eqref{utau} and arrive at $|\Psi_f\rangle$. Finally we apply tomographic pulses and readout the population on $|3\rangle$ by fluorescence measurement. Here as an example the tomographic pulses correspond to measuring Re($\langle 1| \rho_f |4\rangle$). The labels on pulses are explained in the main text.}
	\label{fig01}
\end{figure*}

Secondly,  consider the scenario of two spins $a$ and $b$ in free space, they interact with each other through the magnetic field they produce, i.e., the magnetic dipole-dipole interaction \cite{levitt2013spin,johanning2009quantum}. The derivation of this interaction in QFT framework is given in Sec.\ref{a1} \cite{wang2018magnetic,hu2020field}. When the light crossing time is negligible,  the interaction Hamiltonian is well approximated by
$
H_{dd}=\lambda [3(\mathbf{I}_a \cdot \mathbf{r})(\mathbf{I}_b \cdot \mathbf{r})-\mathbf{I}_a \cdot \mathbf{I}_b],
$
where $\mathbf{I}_{a(b)}$ is angular momentum operator with components $I_{a(b),k}, (k=x,y,z)$, $\mathbf{r}$ is the unit vector in the direction of the line joining the two spins, and $\lambda$ is a constant depending on the distance between the spins  and their gyromagnetic ratios (the tensor products `$\otimes$' between operators are omitted) \cite{levitt2013spin}.
When there is a strong magnetic field present in $z$, $H_{dd}$ could be further simplified to
\begin{equation}\label{sdd}
H_{zz}=\lambda I_{a,z}I_{b,z},
\end{equation}
which is called the secular approximation \cite{levitt2013spin}. This form of interaction will result in a final state similar to \eqref{gf1}, and could be alternatively realized by choosing some proper rotating reference frame, as explained later.
Denote the eigenvalues of $H_{zz}$ to be $S_{nj}$ ($n,j$ could be $\uparrow,\downarrow$), with corresponding eigenstates $|n\rangle|j\rangle$. $S_{nj}$ equals the magnetic potential energy when $a$ is in $|n\rangle$ and $b$ is in $|j\rangle$.
Simulating the GME experiment using magnetic field under $H_{zz}$ is slightly different from the electric field. This is because spin states also determine magnetic field, unlike in GME or the electric field case, where the gravitational field or the electric field directly depends on the position of $a$ and $b$ only. Different spin states $|\psi_a\rangle|\psi_b\rangle$ could give same gravitational field (as long as the position configurations are the same), while this is generally not true for the magnetic field case. Consequently, the protocol using magnetic field is as follows.

Assume at the beginning, $a$ and $b$ are far apart (or magnetically shielded) from each other such that the interaction between them could be ignored, and their spin states are separable.  First prepare each spin to be $\frac{1}{\sqrt{2}}(\left|\uparrow \right\rangle+\left|\downarrow\right\rangle)$.
To see the time-evolved state, first note that
starting from $|n\rangle |j\rangle |M_{0}\rangle$, where $n,j=$ $\uparrow,\downarrow$ and $|M_{0}\rangle$ is the initial magnetic field state, the total state of spin and field at time $t$ under the QFT interaction Hamiltonian $H_{\textrm{F}}$ (containing the spin$+$field degrees of freedom, see Sec.\ref{a1}) could be written as $\alpha(t)|n\rangle |j\rangle |M_{nj}(t)\rangle+\sum_{n'j'}\beta_{n'j'}(t)|n'\rangle |j'\rangle |M_{n'j'}(t)\rangle$, where $n'j'$ are  combinations different from $nj$, $|\alpha(t)|^2+\sum_{n'j'}|\beta_{n'j'}|^2=1$ and  $|M_{nj(n'j')}(t)\rangle$ is the corresponding magnetic field state.
As $H_{\textrm{F}}$  could be simplified to $H_{zz}$, we have $|\alpha(t)| \approx 1$ and $|M_{nj}(t)\rangle \approx |M_{0}\rangle$, that is, the field state remains unchanged approximately and the initial state is an approximate eigenstate of $H_{\textrm{F}}$ hence only generates a phase during the interaction period. On the other hand, the difference of $|M_{nj}(t)\rangle$ for different $nj$, although tiny, is crucial in generating the final entanglement of spins, as explained in Sec.\ref{a1}. So we write $|M_{nj}(t)\rangle=|M_{nj}\rangle$ as a constant state and set $|\alpha(t)|=1$.
Thus after letting the two spins interact through magnetic field for time $\tau$, the final state is
\begin{equation}\label{mf1}
\begin{aligned}
|\Psi_f\rangle=&\frac{1}{2}(e^{-i\phi_{\uparrow \uparrow}}\left|\uparrow\right\rangle\left|\uparrow\right\rangle |M_{\uparrow\uparrow}\rangle+\left|\uparrow\right\rangle\left|\downarrow\right\rangle |M_{\uparrow\downarrow}\rangle \\
&+\left|\downarrow\right\rangle\left|\uparrow\right\rangle |M_{\downarrow\uparrow}\rangle+e^{-i\phi_{\uparrow \uparrow}}\left|\downarrow\right\rangle\left|\downarrow\right\rangle |M_{\downarrow\downarrow}\rangle),
\end{aligned}
\end{equation}
omitting the irrelevant global phase, where
$\phi_{\uparrow \uparrow}=(S_{\uparrow\uparrow}-S_{\uparrow\downarrow})\tau=\lambda\tau/2$.

Finally, we are going to measure and detect entanglement. We choose to perform quantum state tomography, as it has the ability to conclude a definitive negative experimental outcome (i.e., no entanglement), and is also more straightforward to implement on our setup. Unlike the gravitational field case, we are not able to disentangle the magnetic field state. However, 
as explained in Sec.\ref{details} and Sec.\ref{a1}, the fact that $H_F$ could be simplified to $H_{zz}$ indicates $|M_{nj}\rangle \approx |M_{n'j'}\rangle$ in \eqref{mf1}.  Denote $\rho_f$ the density matrix after tracing out the field degrees of freedom in \eqref{mf1}, we could thus apply conventional two-qubit state tomography technique and calculate the entanglement of formation $W(\rho_f)$ \cite{wootters2001entanglement} to detect entanglement.
This also suggests that if the field states in different branches are made close to each other in future GME experiments, then it might not be necessary to coherently bring back the masses (as this might introduce additional experimental error). One could trace out the field and detect entanglement between the spins.

We are going to simulate the GME experiment using an electron spin and a nuclear spin in a single atom, as shown in Fig.\ref{fig01}(b). The idea is essentially the same as explained above.
Consider the s-electron in the outermost shell and the nucleus in a Hydrogen-like atom, or in an alkaline‐like ion, e.g., the $^{171}$Yb$^{+}$ ion used here.  The s-electron forms a spherically symmetric charge distribution surrounding the nucleus. The nuclear spin feels the magnetic field produced by the electron and vice versa. The system could be treated as a uniformly magnetised sphere interacts with a spin at its center \cite{foot2004atomic}. The magnetic dipole-dipole interaction between two dipoles is already derived in sec.\ref{a1}, under QFT framework, and it could be well approximated by $H_{\textrm{dd}}$ when the two dipoles are close to each other. So if we neglect the light crossing time (thus a QFT or a quantum-controlled classical field derivation could be conveniently approximated by a non-relativistic quantum mechanics one), to find the approximation of the QFT description, we could directly starting from $H_{\textrm{dd}}$.  Then following e.g., Ref. \cite{foot2004atomic} we obtain the hyperfine interaction
$
H_h=A(I_{a,x}I_{b,x}+I_{a,y}I_{b,y}+I_{a,z}I_{b,z}),
$
where $A$ is the hyperfine interaction constant, $a$ represents nucleus and $b$ electron. The influence of the remaining symmetrical core of paired electrons could be approximated as a correction term that contributes to the interaction constant \cite{ahmad1982theory}.
The ion is typically in an external magnetic field $B_0$ providing the quantization axis (denote it $z$ axis), so the free evolution Hamiltonian is
\begin{equation}\label{hh}
\begin{aligned}
H_f&=A(I_{a,x}I_{b,x}+I_{a,y}I_{b,y}+I_{a,z}I_{b,z})\\
&-B_0(\gamma_aI_{a,z}I+\gamma_bII_{b,z})\\
&=E_1|1\rangle\langle 1|+E_2|2\rangle\langle 2|+E_3|3\rangle\langle 3|+E_4
|4\rangle\langle 4|,
\end{aligned}
\end{equation}
where $\gamma_{a,b}$ are the gyromagnetic ratios, $I$ is the $2\times2$ identity matrix. $|1\rangle$, $|2\rangle$, $|3\rangle$, $|4\rangle$ and $\left|\uparrow \right\rangle \left|\uparrow\right\rangle$, $\left|\uparrow \right\rangle \left|\downarrow\right\rangle$, $\left|\downarrow \right\rangle \left|\uparrow\right\rangle$, $\left|\downarrow \right\rangle \left|\downarrow\right\rangle$ are related by a mapping operator $R$ as explained in sec.\ref{aexp}:
\begin{equation}
\begin{aligned}
&|1\rangle=\left|\uparrow \right\rangle \left|\uparrow\right\rangle,\quad |2\rangle=\cos\frac{\theta}{2}\left|\uparrow \right\rangle \left|\downarrow\right\rangle-\sin\frac{\theta}{2}\left|\downarrow \right\rangle \left|\uparrow\right\rangle \\
&|3\rangle=\sin\frac{\theta}{2}\left|\uparrow \right\rangle \left|\downarrow\right\rangle+\cos\frac{\theta}{2}\left|\downarrow \right\rangle \left|\uparrow\right\rangle,\quad |4\rangle=\left|\downarrow \right\rangle \left|\downarrow\right\rangle,
\end{aligned}
\end{equation}
where $\theta$ depends on $B_0$ ($\theta \approx -\frac{\pi}{2}$ here).
In $^{171}$Yb$^{+}$, \eqref{hh} corresponds to the $^{2}S_{1/2}$ four-level subspace, as shown in Fig.\ref{fig01}(b).
Although \eqref{hh} is different from \eqref{sdd} in the lab frame, \eqref{hh} seen from a rotating reference frame $f_O$ will take the same form as \eqref{sdd}:
Define a rotating frame $f_O$ with rotation operator $O$, $|\psi\rangle_R=O|\psi\rangle_L$, where $|\psi\rangle_L$ is the state in the lab frame and $|\psi\rangle_R$ in the rotating frame. The evolution operators and Hamiltonian are also changed accordingly. The rotation operator is
\begin{equation}\label{rot}
\begin{aligned}
O=&\textrm{exp}[i\int_{0}^{t'}(\delta_1(t)|1\rangle\langle 1|+\delta_2(t)|2\rangle\langle 2|\\&+\delta_3(t)|3\rangle\langle 3|+\delta_4(t)
|4\rangle\langle 4|)dt].
\end{aligned}
\end{equation}
Working in this frame makes control-sequence-designing easier and it does not affect the population readout.
In the interaction period $\tau$, we set $\delta_{1,4}=E_{1,4}-\Delta$ and $\delta_{2,3}=E_{2,3}$. The evolution operator in this period is then
\begin{equation}\label{utau}
\begin{aligned}
U(\Delta,\tau)&=\textrm{exp}[-i(OH_fO^{\dagger}+i\dot{O}O^{\dagger})\tau]\\
&=\textrm{exp}[-i(\Delta|1\rangle\langle 1|+\Delta|4\rangle\langle 4|)\tau]\\
&=\textrm{exp}[-i(\frac{\Delta}{2}|1\rangle\langle 1|-\frac{\Delta}{2}|2\rangle\langle 2|\\&-\frac{\Delta}{2}|3\rangle\langle 3|+\frac{\Delta}{2}|4\rangle\langle 4|)\tau]\textrm{exp}[-i\frac{\Delta}{2}\tau]\\
&=\textrm{exp}(-i2\Delta I_{a,z}I_{b,z}\tau)\textrm{exp}(-i\frac{\Delta}{2}\tau).
\end{aligned}
\end{equation}
This is the desired evolution under the Hamiltonian of the form $H_{zz}$ (ignoring the irrelevant global phase  $\textrm{exp}(-i\frac{\Delta}{2}\tau)$).
The fact that the last equality holds is crucial as this allows a straightforward implementation of the experiment. We also set $\delta_{1,2,3,4}=E_{1,2,3,4}$ during the initial state preparation and final tomography steps. With appropriate choice of the control-pulse phases,  in this $f_O$, the interaction between spins is ``tailored'' to the form of $H_{zz}$ during the interaction period. The fact that \eqref{rot} is itself an entangling operation does not affect the validity of the simulator [See sec.\ref{details} for explanation].


To sum up, the protocol using a single ion is

1) First prepare the state
\begin{equation}\label{psi0single}
\begin{aligned}
|\Psi_0\rangle=&\frac{1}{2}(\left|\uparrow\right\rangle\left|\uparrow\right\rangle |M_{\uparrow\uparrow}\rangle+\left|\uparrow\right\rangle\left|\downarrow\right\rangle |M_{\uparrow\downarrow}\rangle\\&+\left|\downarrow\right\rangle\left|\uparrow\right\rangle |M_{\downarrow\uparrow}\rangle+\left|\downarrow\right\rangle\left|\downarrow\right\rangle |M_{\downarrow\downarrow}\rangle)
\end{aligned}
\end{equation}
in $f_O$.

2) Then let the system evolves for time $\tau$. In $f_O$, the evolution is  under the Hamiltonian
$
H=2\Delta I_zI_z,
$
where $\Delta$ could be varied by adjusting $\delta_{1,2,3,4}$ in \eqref{rot}, and the final state
\begin{equation}\label{mf2}
\begin{aligned}
|\Psi_f\rangle=&\frac{1}{2}(e^{-i\phi_{\uparrow \uparrow}}\left|\uparrow\right\rangle\left|\uparrow\right\rangle |M_{\uparrow\uparrow}\rangle+\left|\uparrow\right\rangle\left|\downarrow\right\rangle |M_{\uparrow\downarrow}\rangle\\&+\left|\downarrow\right\rangle\left|\uparrow\right\rangle |M_{\downarrow\uparrow}\rangle+e^{-i\phi_{\uparrow \uparrow}}\left|\downarrow\right\rangle\left|\downarrow\right\rangle |M_{\downarrow\downarrow}\rangle),
\end{aligned}
\end{equation}
where
$\phi_{\uparrow \uparrow}=\Delta\tau$ is achieved.

3) Detect entanglement of the two spins in \eqref{mf2} after tracing out the field.




\section{Observation of the field mediated entanglement}\label{ob}

The experiment is performed on the trapped $^{171}$Yb$^{+}$ ion quantum information processor, as shown in Fig.\ref{fig01}(b). A magnetic field with strength $B_0$ along $z$ provides the quantization axis. The $^2$S$_{1/2}$ hyperfine energy levels result from the hyperfine interaction between the $^{171}\textrm{Yb}$ nuclear spin and the s-electron spin in the valence shell.
A microwave field with tunable frequency, amplitude and phase is used to control the spin states. The population on $|3\rangle$, $P_{3}$ could be readout by first measuring the total population $P_1+P_2+P_4$ on $|1\rangle$, $|2\rangle$, and $|4\rangle$ through applying the $369.5$ nm detection beam and measuring the fluorescence, then $P_{3}=1-(P_1+P_2+P_4)$. The initial state of each experiment is $|3\rangle$, which is dissipatively prepared by optical pumping (note interestingly that $|3\rangle$ is actually already an entangled state). Details of the experimental apparatus and parameters of the four-level subspace are given in Sec.\ref{aexp}.

\begin{figure}[htb]  
	\makeatletter
	\def\@captype{figure}
	\makeatother
	\centering
	\includegraphics[scale=0.82
	]{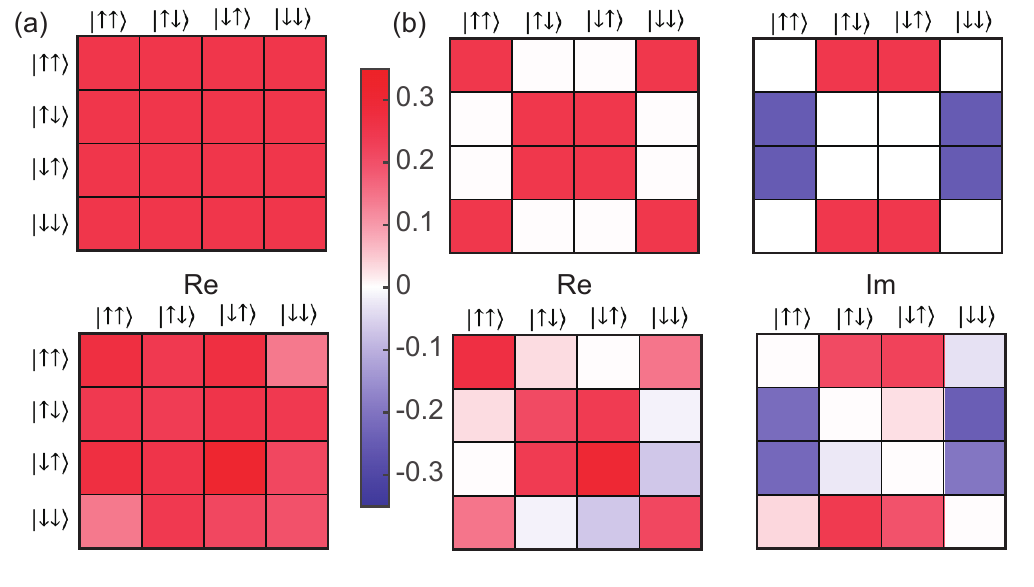}
	\caption{(a) Real parts of the quantum state tomography of  $\rho_{\textrm{exp}}(\phi_{\uparrow\uparrow}=0,\tau=0)$ (lower panel) and the corresponding $\rho_{\textrm{th}}$ (upper panel). The imaginary parts of $\rho_{\textrm{th}}$, $\textrm{Im}(\rho_{\textrm{th}})$ are all zero,  $|\textrm{Im}(\rho_{\textrm{exp}})|$ are all smaller than $0.06$ and are not shown. The fidelity $F \approx 0.97\pm0.01$. $|i\rangle|j\rangle$ are shown as $|ij\rangle$, ($i,j=$ $\uparrow$ or $\downarrow$) for simplicity. (b) Real and imaginary parts of the quantum state tomography of $\rho_{\textrm{exp}}(\phi_{\uparrow \uparrow}=\pi/2,\tau=2$ $\mu$s), $F \approx 0.98\pm0.01$}
	\label{fig04}
\end{figure}

The experimental sequence is illustrated in Fig.\ref{fig01}(c), in the following, the states and operators are all in $f_O$:

1) After initializing the system to $|3\rangle$ through optical pumping, prepare the state $|\Psi_0\rangle$ by sequentially applying  microwave pulses resonant with transitions $|3\rangle \leftrightarrow |4\rangle$, $|1\rangle \leftrightarrow |3\rangle$, $|2\rangle \leftrightarrow |3\rangle$. Here pulses in transitions $|n\rangle \leftrightarrow|m\rangle$ ($n<m$) with labels $\pm y,\theta$ represent evolution operators $U_{p}=\textrm{exp}[\mp i(-i|n\rangle \langle m|+i |m\rangle \langle n|)\theta]$. The desired evolution could be realized by adjusting the pulse width, microwave field strength and phase, as explained in Sec.\ref{aexp}, $\theta_{13,23}$ are also given there.

2) The interaction period. Evolve the system by $U(\Delta,\tau)$ (eq.\eqref{utau}) for $\tau$. To achieve this, recall that we are in $f_O$, we just need to let the system evolve freely for $\tau$. For this to work, we have to apply an extra $\phi=\Delta\tau$ phase  shift for the following readout microwave pulses. This is a common technique in quantum information processing experiments to achieve operations under $H_{zz}$, e.g., when implementing  controlled-phase (CZ) gates \cite{vandersypen2005nmr,haffner2008quantum}.

3) Finally, we apply quantum state tomography to obtain the full density matrix of the two-spin system, and detect entanglement. As the direct observable is $|3\rangle\langle 3|$, we convert the populations and coherence to the $\{|2\rangle,|3\rangle\}$, $\{|1\rangle,|3\rangle\}$ , or $\{|3\rangle,|4\rangle\}$ subspace by $\pi$ pulses, and then read them out by standard qubit-tomography techniques.  For example, in Fig.\ref{fig01}(c) we show the readout pulses obtaining Re($\langle 1| \rho_f |4\rangle$). Assume after this sequence the population on $|3\rangle$ is measured to be $P'_3$,  and we have already measured $P_1$ and $P_4$ (by $\pi$ pulses on $|1,4\rangle \leftrightarrow |3\rangle$ and readout), then $\textrm{Re}(\langle 1| \rho_f |4\rangle)=(P_1+P_4)/2-P'_3$. After obtaining all  $\langle u| \rho_f |v\rangle$, ($u,v=$ $1$, $2$, $3$, $4$), we easily recover $\langle nj| \rho_f |kl\rangle$, $(n,j,k,l=$ $\uparrow$ or $\downarrow$) by applying the mapping operator $R$.

When $\tau=0$, the system is in the initial separable state. The theoretical and experimental density matrix $\rho _{\textrm{th}}$ and $\rho _{\textrm{exp}}$  are shown in Fig.\ref{fig04}(a). The state fidelity
$F=\left| \mathrm{Tr}( \rho_{\textrm{th}}\rho_{\textrm{exp}}) \right|/\sqrt{\mathrm{Tr}(\rho_{\textrm{th}}\rho_{\textrm{th}})\mathrm{Tr}(\rho_{\textrm{exp}} \rho_{\textrm{exp}})}\approx 0.97 \pm 0.01$. To further obtain the degree of entanglement, we calculate the entanglement of formation $W(\rho)$  of $\rho$. $W[\rho_{\textrm{exp}}(\phi_{\uparrow \uparrow}=0)]=0$, confirming the separability.

After an interaction period $\tau$, the two spins are generally entangled. For example, we choose $\tau=2$ $\mu$s, $\Delta=\pi/(2\tau)=0.785$ MHz, and the final state will be maximally entangled ($\phi_{\uparrow \uparrow}=\pi/2$). This could serve to simulate a GME experiment with same $\phi_{\uparrow \uparrow}$, e.g, $m=10^{-14}$ kg, $\tau=2.5$ s, $d_{uu}=200$ $\mu$m, $d_{ud}=280$ $\mu$m, which could be realized by future interferometers using masses with embedded spins \cite{bose2017spin}. The experimental results are shown in Fig.\ref{fig04}(b), where $F\approx0.98\pm 0.01$.
$W[\rho_{\textrm{exp}}(\phi_{\uparrow \uparrow}=\pi/2)]=0.66\pm0.06>0$, which means the state is entangled. Deviation from the ideal value $1$ is mainly due to pulse errors and decoherence in steps 1) and 3).
We also vary $\Delta$ with fixed $\tau=2$ $\mu$s, the results are, e.g., $W[\rho_{\textrm{exp}}(\phi_{\uparrow \uparrow}=\pi/4)]=0.40\pm0.02$ with $F=0.98\pm0.01$ ($W[\rho_{\textrm{th}}(\phi_{\uparrow \uparrow}=\pi/4)]=0.5$), and $W[\rho_{\textrm{exp}}(\phi_{\uparrow \uparrow}=\pi)]=0.02^{+0.03}_{-0.02}$ with $F=0.98\pm0.01$ ($W[\rho_{\textrm{th}}(\phi_{\uparrow \uparrow}=\pi)]=0$).

\begin{figure}[h]  
	\makeatletter
	\def\@captype{figure}
	\makeatother
	\centering
	\includegraphics[scale=1.0
	]{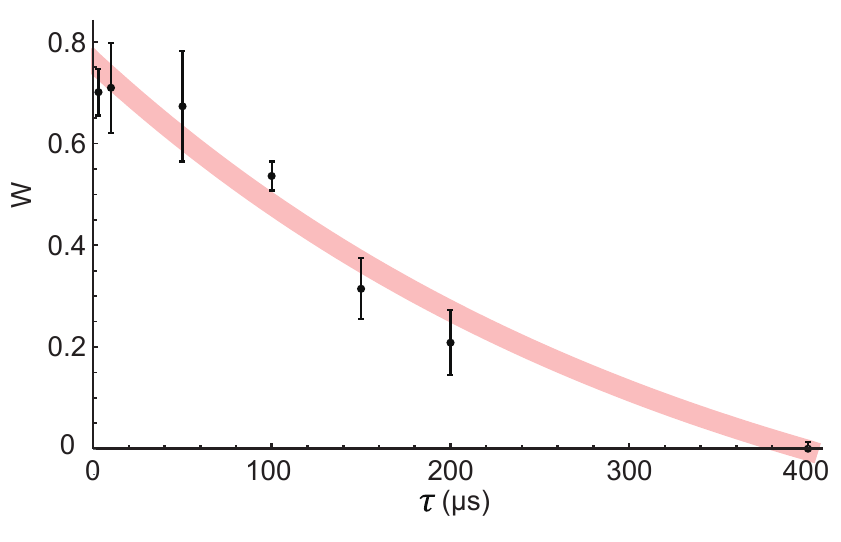}
	\caption{Decoherence of spins due to magnentic noise. The entanglement of formation
		 $W[\rho_{\textrm{exp}}(\phi_{\uparrow \uparrow}=\Delta\tau=\pi/2)]$ is plotted with varying $\tau$ and $\Delta$. The light red curve is added to guide the eye. }
	\label{fig05}
\end{figure}

If a negative outcome is encountered in future GME experiments, it could results from the following three possibilities: 1. Gravity does not have standard quantum mechanical properties; 2. There are experimental noises such as magnetic noise and gravitational noise that causes decoherence; 3. There are other new mechanisms present, such as the spontaneous collapse mechanism \cite{bassi2013models} that leads to a strong loss of coherence on the time scale of the experiment.
We observe the decoherence effect in the simulation experiment. While in our case such an effect is due to the magnetic noise described in possibility $2$, it also provides a mechanism to simulate possibility $3$. In the $zz$ basis (defined in Sec.\ref{aexp}),
denote the density matrix corresponds to the spin part of \eqref{gf1} $\rho_f(\tau)=1/4\sum_{n,j,k,l} \alpha_{njkl}|nj\rangle\langle kl|$, ignoring the gravitational part as it is already factored out. Assume for simplicity that  the decoherence introduces a decay factor $e^{-\tau/\beta_{njkl}}$ to each nondiagonal element, then $\alpha_{njkl}=e^{-\tau/\beta_{njkl}}e^{i\varphi_{njkl}}$, where $e^{i\varphi_{njkl}}$ are phases without decoherence. Denote the final state of the simulator $\rho^{s}_f(\tau)=1/4\sum_{n,j,k,l} \alpha^{s}_{njkl}|nj\rangle\langle kl|$, also ignoring the magnetic field states. The simulator is also undergoing decoherence process, as   shown by the Ramsey $T^{*}_2$ measurement among different transitions. Through the mapping operator $R$, the exponential decay of the coherence $\langle b|\rho^{s}_f(\tau)|a\rangle$, ($a,b=$ $1$, $2$, $3$, $4$) manifest themselves as the decay factors $e^{-\tau/\beta^s_{njkl}}$ in $\alpha^{s}_{njkl}$. Thus $\alpha^s_{njkl}=e^{-\tau/\beta^s_{njkl}}e^{i\varphi^s_{njkl}}$, where $e^{i\varphi^s_{njkl}}$ are phases without decoherence. Clearly the decoherence in GME could be simulated by the decoherence process in the simulator.
We fix $\phi_{\uparrow \uparrow}=\pi/2$, and vary $\tau$ from $0$ to $400$ $\mu$s ($\Delta$ is varied accordingly). The resulting $W[\rho_{\textrm{exp}}(\phi_{\uparrow \uparrow}=\Delta\tau=\pi/2)]$ is illustrated in Fig.\ref{fig05}. When $\tau$ is small, $W$ is relatively large, indicating an entangled state with high-degree of entanglement.  The deviation of $W$ from ideal value $1$ is due to the error and decoherence in the preparation and tomography steps. As $\tau$ is increased, $W$ decreases, and eventually approaches $0$. This demonstrates the importance of minimizing decoherence due to technical noise in future GME experiments. It would also be crucial to have a good characterization of the magnetic noise if one wishes to further tell apart   decoherence due to spontaneous collapse from decoherence due to magnetic noise.

\section{Discussion}

\subsection{Indistinguishability of the quantum-controlled classical field and the QFT description}
The result of entanglement detection in this work ``rules out'' mean-field source for the electromagnetic potential  \cite{fragkos2022inference}. As described in Sec.\ref{a1}, the result is in consistent with both the QFT description of electromagnetic field and the quantum-controlled classical field description. So, the distinguishability of the two scenarios is lacking in the current GME experiments. To further tell these two descriptions apart,  measuring on the order of the light-crossing time in future GME experiments is required, that is, demonstrating the two masses begin to get entangled due to gravitational interaction within the light crossing time, or measuring tinier quantitative effects around the light-crossing time \cite{martin2022gravity}.
It is interesting to point out that existing experiments on  entanglement between remote trapped-ion qubits \cite{moehring2007entanglement,luo2009protocols,hannegan2022entanglement} could be modified to a demonstration of this effect in electromagnetism.
For example, in Ref.\cite{moehring2007entanglement}, starting from time $t=0$,
two $^{171}$Yb$^{+}$ atomic qubits ($|2,3\rangle$ form the qubit levels) remotely located with light crossing time $t_c$ get entangled  mediated by electromagnetic field, i.e., the photons they produce. To achieve this, each atom first emit a single photon to create an entangled state between the atomic qubit and the photonic qubit (frequency qubit). The photons are then directed to a beam splitter to perform two-photon-interference. Coincident photon detection at time $t_d$ in the two output ports of the beam splitter projects the atomic qubits to an entangled state. The desired observable of atomic qubits is then measured by internal-state control and readout with a total duration $t_r$.  If the experimental setup could be modified such that $t_d+t_r<t_c$, one would be able to show that the entanglement is established within the light crossing time between the two atomic qubits, thus falsify the quantum-controlled classical field description. The implication of this  for GME is worth further study.

\subsection{Two-ion entanglling gate as a simulation of  GME}\label{a21}
Electromagnetic field's ability to stay in quantum superposition correlated with the source is necessary for two-qubit entanglling gates based on Coulomb interaction in trapped ions, e.g., the Cirac-Zoller gate \cite{cirac1995quantum}, M\o lmer-S\o rensen gate \cite{sorensen1999quantum,kirchmair2009deterministic}, light-shift gate \cite{leibfried2003experimental}, and fast gates that are not confined within the Lamb-Dicke regime \cite{wong2017demonstration}.  Starting from $1/2(\left|\uparrow \uparrow\right\rangle+\left|\uparrow \downarrow\right\rangle+\left|\downarrow \uparrow\right\rangle+\left|\downarrow \downarrow\right\rangle)$, during these gates, entangled states between internal  degrees of freedom (spin) and the collective motional degrees of freedom (phonon) are generated. After motional states in all branches (correspond to different collective spin states)  become the same and factored out, an entangled state of two spins is generated. Relative phases among different branches are coherently accumulated
under the joint action of spin energy, motional energy, interaction energy with the control field, and importantly, the electromagnetic field energy caused by the electromagnetic interaction between two ions. If there is no such interaction, the center of mass mode and relative motion mode will have same mode frequency, then $\phi_{\uparrow\uparrow}+\phi_{\downarrow\downarrow}=\phi_{\uparrow\downarrow}+\phi_{\downarrow\uparrow}$, hence no entanglement between the two spins will be generated. If the electromagnetic field is not quantum, instead it admits a mean-field description and takes the same state in all branches, then the situation is the same as if there is no electromagnetic interaction between the two ions (plus a global electric field which only creates a global phase), hence no entanglement is generated either.

Take the experiment in Ref.\cite{wong2017demonstration} as an example, two trapped $^{171}$Yb$^{+}$ ions ($|2,3\rangle$ form the qubit levels) get entangled through mutual electrical interaction and interaction with laser control field. One basic operation of the experiment is the ``spin-dependent-kick'' (SDK) that creates entangled states of spin and motion of each ion, which is similar to Stern-Gerlach operation. The SDK evolution operator for the two-qubit system in the rotating frame is $U_{\textrm{SDK}}=e^{2i\phi(t)}\sigma_{1+}\sigma_{2+}D_{\textrm{C}}(i\eta_{\textrm{C}})+e^{-2i\phi(t)}\sigma_{1-}\sigma_{2-}D_{\textrm{C}}(-i\eta_{\textrm{C}})+\sigma_{1+}\sigma_{2-}D_{\textrm{R}}(i\eta_{\textrm{R}})+\sigma_{1-}\sigma_{2+}D_{\textrm{R}}(-i\eta_{\textrm{R}})$, where $\sigma_{1,2+(-)}=\left|\downarrow\right\rangle \left\langle\uparrow\right|(\left|\uparrow\right\rangle \left\langle\downarrow\right|)$ for spin $1$ and $2$, $\phi(t)$ and $\Delta k$ is related to the laser control field, $x_{1,2}$ are position operators for ion $1$ and $2$,  $D_{\textrm{C},\textrm{R}}(i\eta_{\textrm{C},\textrm{R}})$ are coherent state displacement operators for the center of mass mode and relative motion mode, and $\eta_{C,R}$ are the corresponding Lamb-Dicke parameters.  Starting from $1/2(\left|\uparrow \uparrow\right\rangle+\left|\uparrow \downarrow\right\rangle+\left|\downarrow \uparrow\right\rangle+\left|\downarrow \downarrow\right\rangle)$, combination of several $U_{\textrm{SDK}}$ and free evolution disentangles the motional state and generates phase difference among different branches. The final entangled state is
$|\Psi_e\rangle=e^{i\Phi_g}/2(e^{i\gamma}\left|\downarrow\right\rangle\left|\downarrow\right\rangle+e^{-i\gamma}\left|\uparrow\right\rangle\left|\uparrow\right\rangle)-1/2(\left|\uparrow\right\rangle\left|\downarrow\right\rangle+\left|\downarrow\right\rangle\left|\uparrow\right\rangle)$, where $\Phi_g=\pi/2.06$ and $\gamma$ is a constant in each experiment. The difference of electromagnetic field in different branches is crucial to generate $\Phi_g \neq n\pi$ ($n$ is an integer), i.e., entanglement.
As a simulation of GME, the electromagnetic field plays the role of gravitational field. Different  motional states of the two ions play the role of different positional combination of the masses. The combination of $U_{\textrm{SDK}}$ and free evolution is the anolog of the interaction period (for $\tau$) plus ``bringing back to the middle position'' in GME.

\section{Conclusion}
To conclude, by taking the QFT of electromagnetic field as a premise, this work propose and perform a quantum simulation experiment of the GME effect. It could provide a preview of the general experimental procedure of the future GME experiment.
 Observing that the experimental result could not distinguish between a QFT description from a quantum-controlled classical field one,  this work helps to stress the benefits of measuring on the order of the light-crossing time.
It also points out that existing experiments on entanglement between remote trapped-ion qubits \cite{moehring2007entanglement,luo2009protocols,hannegan2022entanglement} could be modified to demonstrate such measurements in electromagnetism.

The experimental system consists of two spins and the magnetic field they create which mediates their interaction, in a single atom. The use of the naturally-existing multilevel hyperfine structure in a single ion reduces the experimental cost of simulating the GME experiment. The quantum control techniques presented here could be further developed to achieve experimental complete-control of qudit systems in $^{171}$Yb$^{+}$. This work is also among the first to abstract a naturally-existing two-spin system from a single $^{171}\textrm{Yb}^{+}$ ion \cite{yang2022realizing}   and perform two-qubit simulation experiments upon it, which may shed new light on quantum simulation using trapped ions \cite{monroe2021programmable}.

\section{Appendix}
\subsection{Brief description of GME.}\label{m1}
 Consider two objects $a$ and $b$, each has mass $m$ and posses spin-$1/2$ , as shown in Fig.\ref{fig01}(a).
The radius of the masses are neglected, $d_{uu}=d_{dd}=d_0$, $d_{ud}=d_{du}>d_0$ and the spin-spin interaction between $a$ and $b$ is so weak that could be neglected. Initially one prepares the spin state of each mass to be $\frac{1}{\sqrt{2}}(\left|\uparrow \right\rangle+\left|\downarrow \right\rangle)$ and hold the position of $a$ and $b$ to be in the middle of $u$ and $d$. Denote the state of the gravitational field corresponds to this configuration $\left|g_0 \right\rangle$, then the total system becomes
$
\frac{1}{2}(\left|\uparrow \right\rangle+\left|\downarrow \right\rangle)(\left|\uparrow \right\rangle+\left|\downarrow \right\rangle)|g_{0}\rangle.
$
Next, one releases $a$ and $b$. If there is only one mass, the mass would be in spatial superposition of $u$ and $d$ (neglecting the time it takes to develop this superposition).
As explained in Sec.\ref{realization}, after the gravitational interaction for time $\tau$ the state becomes \eqref{eq2}.
Each component of \eqref{eq2} is approximately an eigenstate of the total Hamiltonian hence only generates a phase during the interaction period \cite{christodoulou2019possibility}.
The phase acquired by the internal (spin) energy $\hbar\omega_{nj}$ $(n,j=$ $ \uparrow \textrm{or} \downarrow)$ is ignored, as one could compensate this by going to the rotating frame.
As $\Phi_{uu}=\Phi_{dd}$, $\Phi_{du}=\Phi_{ud}$, we could  subtract an irrelevant global phase factor and define the new relative phases before each branch as
$\phi_{\uparrow \uparrow}=(\Phi_{uu}-\Phi_{ud})\tau/\hbar,
\phi_{\uparrow \downarrow}=0,
\phi_{\downarrow \uparrow}=0,
\phi_{\downarrow \downarrow}=(\Phi_{uu}-\Phi_{ud})\tau/\hbar.$ Thus one obtains the final state \eqref{gf1}.
In the full language of general relativity, there are higher-order corrections to the phases. For example, consider the leading correction term to $\phi'_{\uparrow\uparrow}$, $\tau/\hbar E_{\textrm{m}}=\tau/\hbar[\Phi_{uu}\hbar(\omega_{\uparrow}+\omega_{\uparrow})/(mc^2)]$, where $E_{\textrm{m}}$ could be intuitively think of as originated from the  internal energy contribution to the weight of the system and thus also couples to gravity \cite{pikovski2015universal}. These terms are neglected in the original GME proposals, as $\hbar(\omega_{n}+\omega_{j})/(mc^2) \ll 1$ for typical experimental parameters. Interestingly, it is argued that measuring these terms would allow one take a step forward  from concluding the superposition of gravitational fields to concluding the superposition of spacetime \cite{qiss,christodoulou2019possibility}, as these terms correspond to the gravitational time dilation (red shift) and could be
measured by quantum clocks \cite{chou2010optical,bothwell2022resolving}.

\subsection{Details of the single-atom GME simulator}\label{details}
We first explain entanglement detection in \eqref{mf1}. Unlike the gravitational field case, one is not able to disentangle the magnetic field state. Nevertheless, as is derived in Sec.\ref{a1}, the QFT Hamiltonian $H_F$ is well approximated by $H_{dd}$, this means the electromagnetic field state is almost unchanged, as $H_{dd}$ only affects the spins, not the electromagnetic field. Starting from $H_{dd}$, one obtains $H_h$, and hence $H_{zz}$ in the simulator. This results in $|M_{\uparrow\uparrow}\rangle \approx |M_{\uparrow\downarrow}\rangle \approx |M_{\downarrow\uparrow}\rangle \approx |M_{\downarrow\downarrow}\rangle\approx|M_{0}\rangle$. So tracing out the field state in \eqref{mf2}, one obtains a density matrix that is negligibly different from the density matrix after disentangling the field. Note however that the tiny difference between field states in different branches is required to generate different phases, hence the final entanglement, as explained in Sec.\ref{a1}.
To further obtain the degree of entanglement, for a two-qubit state $\rho$ (pure or mixed), one could calculate its entanglement of formation $W(\rho)$ \cite{wootters2001entanglement}. A larger $W$ indicates a higher degree of entanglement: $W=0$ indicates a separable state while $W=1$ a maximally entangled state.

We then explain the rotating operator \eqref{rot}. Observing that \eqref{rot} is an entangling operation might raises concern: whether the entanglement of the spins is due to the quantum properties of the magnetic field, or is merely an artifact due to the rotating frame transformation. This is not a problem, as already mentioned that the logic of this work is to take the quantum nature of electromagnetic fields as premise, and conduct a simulation experiment that might bring insights to the GME experiment. Transforming to the rotating frame merely makes the simulation more close to the original GME experiment, i.e., their final entangled states have the same form. We could have stayed in the lab frame and utilize the non-separable evolution under $H_f$,  to achieve an entangled spin state, but the form of the entangled state will not be the same as \eqref{mf1}. The introduction of $f_O$ merely ``tailored'' the naturally-existing entangling-interaction between spins to the form of $H_{zz}$ during the interaction period.
\subsection{Experimental setup}\label{aexp}

The magnetic field $B_0$ defining the quantization axis ($z$ axis) is created by the electric current in a Helmholtz coil. The electric current is produced by a dc power supply, and is stabilized at $1.20152 \pm 0.00002$ A by a proportional-integral (PI) controller. This allows us determine the resonance frequencies of all three transitions (by Ramsey curves) within $3$ kHz during data collection. Hence we obtain the frequencies of the on-resonance microwaves to be applied (Table.\ref{tab1}), and $B_0$ (used to design the pulse sequence and calculate $R$) is calculated to be $5.615$ Gs by the transition $|3\rangle \leftrightarrow |1\rangle$.
From simulation results, the final state fidelity corresponds to, e.g., $\rho(\phi_{\uparrow \uparrow}=\pi/2,\tau=2$ $\mu$s) is still above $0.97$ when there is $3$ kHz mismatch in all three resonance frequencies ($|3\rangle \leftrightarrow |4\rangle$, $|3\rangle \leftrightarrow |1\rangle$, $|3\rangle \leftrightarrow |2\rangle$), plus $2$ mGs miscalibration (corresponds to the $3$ kHz mismatch) in the magnetic field. This allows us achieve high-fidelity experimental results.

\begin{figure*}[htb]  
	\makeatletter
	\def\@captype{figure}
	\makeatother
	\includegraphics[scale=0.9
	]{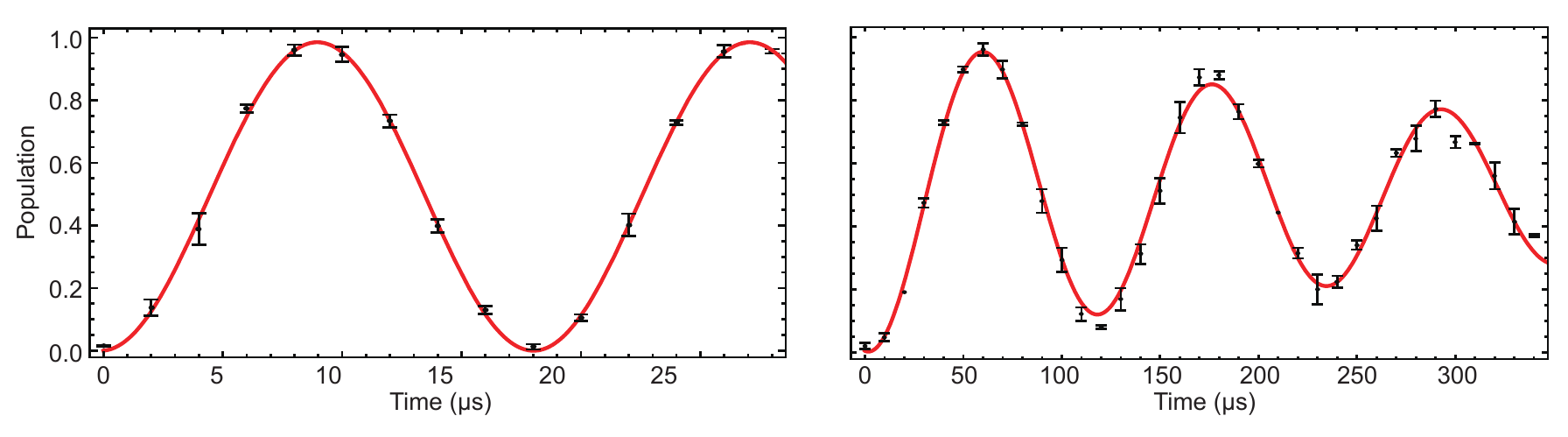}
	\caption{(a) Rabi curve of $|3\rangle\leftrightarrow|4\rangle$. $\Omega/(2\pi)$ is fitted to be $55.2$ kHz. (b) Ramsey measurement of $T^{*}_2$ of $|2\rangle\leftrightarrow|4\rangle$. $T^{*}_2$ is fitted to be $440.7$ $\mu$s. }
	\label{a01}
\end{figure*}


The frequencies of the applied on-resonance microwaves $\omega=\omega_0+\omega'$ ($\omega_0=2\pi\times12.61157173$ GHz), amplitude of Rabi frequencies $|\Omega|$ and coherence times $T^{*}_2$ of the relevant transitions within the four-level subspace (Fig.\ref{fig01}(b)) are given in Table.\ref{tab1}. As an example, the Rabi curve of $|3\rangle\leftrightarrow|4\rangle$ is shown in Fig.\ref{a01}(a), and the Ramsey measurement of $T^{*}_2$ of  $|2\rangle\leftrightarrow|4\rangle$ is shown in Fig.\ref{a01}(b). The $T^{*}_2$ between different Zeeman sublevels could be easily upgraded to above $1$ ms using triggering to ac-line techniques \cite{ruster2016long}.

Details of the quantum-control technique for the four-level/two-qubit system is given below. We first introduce a mapping operator that simplifies the discussion. Define the ``zz basis'' where $\left|\uparrow \right\rangle \left|\uparrow \right\rangle_{\textrm{zz}}=[1,0,0,0]^{\textrm{T}}$, $\left|\uparrow \right\rangle \left|\downarrow \right\rangle_{\textrm{zz}}=[0,1,0,0]^{\textrm{T}}$, $\left|\downarrow \right\rangle \left|\uparrow \right\rangle_{\textrm{zz}}=[0,0,1,0]^{\textrm{T}}$, $\left|\downarrow \right\rangle \left|\downarrow \right\rangle_{\textrm{zz}}=[0,0,0,1]^{\textrm{T}}$, the subscript indicates zz basis.
Also define the ``number basis'' where $|1\rangle_{\textrm{n}}=[1,0,0,0]^{\textrm{T}}$, $|2\rangle_{\textrm{n}}=[0,1,0,0]^{\textrm{T}}$, $|3\rangle_{\textrm{n}}=[0,0,1,0]^{\textrm{T}}$, $|4\rangle_{\textrm{n}}=[0,0,0,1]^{\textrm{T}}$,
the subscript ``n'' indicates number basis.
Define a mapping operator $R$ (depends on $B_0$) satisfying
\begin{equation}
\begin{aligned}
R|k\rangle_{\textrm{n}}=|k\rangle_{\textrm{zz}} \quad (k=1,2,3,4),
\end{aligned}
\end{equation}
i.e., it maps the matrix form of an operator in number basis to that in zz basis, and vice versa.
The explicit form of $R$ reads
\begin{equation}\label{mapr}
R=
\begin{pmatrix} 1 & 0 & 0 & 0 \cr 0 & \cos(\theta/2) & \sin(\theta/2) & 0 \cr 0 & -\sin(\theta/2) & \cos(\theta/2) & 0 \cr 0 & 0 & 0 & 1 \end{pmatrix},
\end{equation}
where $\theta=2 \textrm{tan}^{-1}(\lambda)$ and
\begin{equation}
\lambda=\frac{-B_0\gamma_a+B_0\gamma_b-\sqrt{A^2+B^2_0\gamma^2_a+B^2_0\gamma^2_b-2B^2_{0}\gamma_a\gamma_b}}{A}.
\end{equation}
When $A \gg B_0\gamma_{a,b}$, as in our case, $\theta \approx -\pi/2$. Nevertheless we still use the explicit form of \eqref{mapr}.
During step 1) in the experiment, the Hamiltonian in $f_O$ after rotating wave approximation (RWA) reads
\begin{equation}\label{hcrwa}
H_c=c_{13}(t)|1\rangle\langle 3|+c_{23}(t)|2\rangle\langle 3|+c_{34}(t)|3\rangle\langle 4|+\textrm{H.c.},
\end{equation}
where $c_{13,23,34}$ are complex numbers determined by the strength and phase of the on-resonance microwave control fields.  We tune the microwave field parameters to realize $U_p$ in step $1$ of the experiment. $\theta_{13}=2 \sin^{-1}(1/\sqrt{3})$, and
\begin{equation}
\begin{aligned}
\theta_{23}=&2 \cos^{-1}\{\frac{-\sqrt{2}/2[1-\tan^{-1}(\theta/2)]}{\cos(\theta/2)\tan^{-1}(\theta/2)+\sin(\theta/2)}\}.
\end{aligned}
\end{equation}
Step 2) is realized by free evolution plus corresponding phase correction in the following microwave 	readout pulses, as explained in Sec.\ref{ob}.

With $H_f$ and the available control fields, the four-level system is completely controllable, as could be proved following Ref.\cite{albertini2002lie}.  This means arbitrary $\textrm{SU}(4)$ quantum operations could be achieved in this subspace. In this work, we only pick out several special pieces from the complete control-toolbox, namely, a state-to-state operation achieving \eqref{psi0single}, $\pi$ and $\pi/2$ rotations among the available transitions, and the entangling operation \eqref{utau}. It is promising to demonstrate complete controllability in this system by realizing a universal gate set, i.e., arbitrary single-spin rotations and CNOT gates via, e.g., numerical optimal control \cite{khaneja2005optimal}. Or in the language of qudit, realizing arbitrary single-qudit rotation gates \cite{wang2020qudits,ringbauer2022universal}. The fidelities of the controls and the coherence time could be further upgraded by magnetic shielding \cite{ruster2016long}.

\begin{table}\label{tab1}
	\centering
	\caption{Parameters of the four-level subspace.}
	\begin{tabular}{ccccccc}
		\hline \hline
		& $\omega'/(2\pi)$ $(\textrm{MHz}) \;$ & $|\Omega|/(2\pi)$ $(\textrm{kHz}) \;$ & $T^{*}_2$ $(\mu s)$    \\  \hline
		$|1\rangle\leftrightarrow|3\rangle$                                                               & $39.1096$      & $49.6$ & $404.3$ &    \\
		$|2\rangle\leftrightarrow|3\rangle$                                                  & $31.2500$   & $107.8$  & $>5\times10^{5}$ &  \\
		$|3\rangle\leftrightarrow|4\rangle$                                                         & $23.3814$     & $55.2$  & $403.9$  &     \\
		$|1\rangle\leftrightarrow|2\rangle$                                                     &   /       & / & $410.9$ &    \\
		$|2\rangle\leftrightarrow|4\rangle$                                                    &   /       & / & $440.7$ &        \\
		$|1\rangle\leftrightarrow|4\rangle$                                                     &   /      &  / &  $138.5$ &
		\\ \hline \hline
	\end{tabular}\label{tab1}
\end{table}

\subsection{Magnetic dipole-dipole interaction mediated by electromagnetic field}\label{a1}
Following Ref. \cite{wang2018magnetic,hu2020field}, when the two magnetic dipoles (spin-$1/2$) $\hat{m}_i=\sum_{x,y}\vec{m}^{xy}_i\hat{\tau}^{xy}_i$ ($x,y= \uparrow \textrm{or} \downarrow$, $i=1,2$), with the dipole moment amplitude $\vec{m}^{xy}_i:=\langle x |\hat{m}_i|y\rangle_i$, and $\hat{\tau}^{xy}_i:=|x\rangle_i \langle y|$, are placed in the EM field, they change the original field dynamics. This influence generates the interaction between the two dipoles.  In the Heisenberg picture, the quantized magnetic field in some volume $V
$ is
\begin{equation}
\hat{\mathbf{B}}(\mathbf{x},t)=\sum_{\mathbf{k}\sigma}i\hat{e}_{\mathbf{k}\tilde{\sigma}}Z_{\mathbf{k}}[\hat{a}_{\mathbf{k}\sigma}(t)e^{i\mathbf{k}\cdot\mathbf{x}}-\textrm{H.c.}],
\end{equation}
where $Z_k=\sqrt{\mu_0 \omega_{\mathbf{k}}/(2V)}$, $\hat{e}_{\mathbf{k}\tilde{\sigma}}:=\hat{e}_{\mathbf{k}}\times \hat{e}_{\mathbf{k}\sigma}$ are polarization directions.
 It interacts with the dipole through
the interaction Hamiltonian $-\hat{m}_i \cdot \hat{\mathbf{B}}(\mathbf{x},t)$. The part of the field resulting from the presence of dipole $1$  reads
\begin{equation}\label{bd1}
\begin{aligned}
\hat{\mathbf{B}}_{D1}=&\sum_{\mathbf{k}\sigma,xy}\hat{e}_{\mathbf{k}\tilde{\sigma}}Z_{\mathbf{k}}(g^{xy}_{1,\mathbf{k}\sigma})^{*}\\
&\times e^{i\mathbf{k}\cdot\mathbf{x}}\int_{0}^{t}dse^{-i\omega_\mathbf{k}s}\hat{\tau}^{yx}_1(t-s)+\textrm{H.c.},
\end{aligned}
\end{equation}
where $g^{xy}_{i,\mathbf{k}\sigma}=-i(\vec{m}^{xy}_i \cdot \hat{e}_{\mathbf{k}\tilde{\sigma}})Z_ke^{i\mathbf{k}\cdot \mathbf{x}_i}$. Note $\hat{\tau}^{yx}_1(t-s)$ are operators in Heisenberg picture, they evolve in time (through the Heisenberg equation) and in general act non-trivially on both spin and field states. Its interaction with dipole $2$ results in
\begin{equation}\label{intham}
\begin{aligned}
H_{1\rightarrow 2}&=-\hat{m}_2(t)\cdot \hat{\mathbf{B}}_{D1}(\mathbf{x}_2,t)\\
&=\sum_{uv,xy}\int_{0}^{t}dsD^{yx,uv}_{1\rightarrow 2}(s)\hat{\tau}^{yx}_1(t-s)\cdot\hat{\tau}^{uv}_2(t)+\textrm{H.c.},
\end{aligned}
\end{equation}
where
\begin{equation}
\begin{aligned}
D^{yx,uv}_{1\rightarrow 2}(s):=\sum_{\mathbf{k}\sigma}-i(g^{xy}_{1,\mathbf{k}\sigma})^{*}g^{uv}_{2,\mathbf{k}\sigma}e^{-i\omega_{\mathbf{k}}s}\\
:=-i\int_{0}^{\infty}\frac{d\omega}{2\pi}J^{yx,uv}_{1\rightarrow 2}(\omega)e^{-i\omega s},
\end{aligned}
\end{equation}
and
\begin{equation}
\begin{aligned}
J^{yx,uv}_{1\rightarrow 2}(\omega):=2\pi\sum_{\mathbf{k}\sigma}(g^{xy}_{1,\mathbf{k}\sigma})^{*}g^{uv}_{2,\mathbf{k}\sigma}\delta(\omega-\omega_{\mathbf{k}}).
\end{aligned}
\end{equation}
For free-space EM field,
\begin{equation}
\begin{aligned}
J^{yx,uv}_{1\rightarrow 2}(\omega)=&\frac{\mu_0}{2\pi r^3}\{\vec{m}^{yx}_1 \cdot \vec{m}^{uv}_2[\eta^2\sin\eta+\eta\cos\eta-\sin\eta]\\
&-3(\vec{m}^{yx}_1\cdot\hat{e}_{\mathbf{r}})(\vec{m}^{uv}_2\cdot\hat{e}_{\mathbf{r}})\\
&\times[\frac{1}{3}\eta^2\sin{\eta}+\eta\cos{\eta}-\sin{\eta}]\},
\end{aligned}
\end{equation}
where $\eta:=kr=\omega r/c$, and $r=|\mathbf{x}_1-\mathbf{x}_2|$ is the distance between the two dipoles. The interaction \eqref{intham} has a retarded form, and $D^{yx,uv}_{1 \rightarrow 2}(s)$ characterizes the propagation of the interaction. One could similarly obtain $H_{2\rightarrow 1}$, hence the QFT-interaction Hamiltonian $H_{\textrm{F}}=1/2(H_{1\rightarrow 2}+H_{2\rightarrow 1})$. From the above derivation, we see that $H_{\textrm{F}}$ in general entangles the spins with field. The field is perturbed and deviates from the initial state during the interaction, as could be seen from \eqref{bd1}.

Because $D^{yx,uv}_{1 \rightarrow 2}(s)$ approaches zero rapidly with increasing $s$ \cite{hu2020field}, the interaction is approximately local in time. As $D^{yx,uv}_{1 \rightarrow 2}(s)$ is of the second order of $g^{xy}_{i,\mathbf{k}\sigma}$, we could  make the approximation
\begin{equation}
\hat{\tau}^{yx}_1(t-s)\approx \hat{\tau}^{yx}_1(t)\cdot e^{i\Omega^{yx}_1s},
\end{equation}
where $\Omega^{yx}_1:=E^{(y)}_1-E^{(x)}_1$ is the energy difference between the two levels $|y,x\rangle_1$, hence the interaction becomes time-local.  This approximation means we could neglect the higher-order interaction between the dipole and field. 
As $r/c$ is small, the time to build up the dipole-dipole interaction is small  compared to the interaction period in the experiment. Thus we could extend the time integral to infinity (Markovian approximation), the interaction reduces to a time-independent form
\begin{equation}
H_{1 \rightarrow 2}=\sum_{xy,uv}[K^{yx,uv}_{1\rightarrow 2}(\Omega^{yx}_1)-\frac{i}{2}J^{yx,uv}_{1\rightarrow 2}(\Omega^{yx}_1)]\hat{\tau}^{yx}_1\hat{\tau}^{uv}_2,
\end{equation}
where
\begin{equation}
\begin{aligned}
&K^{yx,uv}_{1\rightarrow 2}(\Omega)=\frac{\mu_0}{4\pi r^3}\{\vec{m}^{yx}_1 \cdot \vec{m}^{uv}_2\\
&[-\eta'^2\cos\eta'+\eta'\sin\eta'+\cos\eta']
-3(\vec{m}^{yx}_1\cdot\hat{e}_{\mathbf{r}})(\vec{m}^{uv}_2\cdot\hat{e}_{\mathbf{r}})\\
&\times[-\frac{1}{3}\eta^2\cos{\eta'}+\eta'\sin{\eta'}+\cos{\eta'}]\},
\end{aligned}
\end{equation}
and $\eta'=\Omega r/c$.
$H_{2 \rightarrow 1}$ could be similarly simplified. The total interaction is then
\begin{equation}\label{hfa}
H'_{F}=\sum_{xy,uv}(G^{(P)}_{yx,uv}+G^{(D)}_{yx,uv})\hat{\tau}^{yx}_1\hat{\tau}^{uv}_2,
\end{equation}
where
\begin{equation}
G^{(P)}_{yx,uv}=\frac{1}{2}[K^{yx,uv}_{1\rightarrow 2}(\Omega^{yx}_1)+K^{uv,yx}_{2\rightarrow 1}(\Omega^{uv}_2)],
\end{equation}
and
\begin{equation}
G^{(D)}_{yx,uv}=\frac{1}{4i}[J^{yx,uv}_{1\rightarrow 2}(\Omega^{yx}_1)+J^{uv,yx}_{2\rightarrow 1}(\Omega^{uv}_2)].
\end{equation}
From the simplified \eqref{hfa}, we see that the field state is almost unperturbed.
In our experiment,   $\Omega^{yx}_{1,2}r/c \rightarrow 0$, then all $G^{(D)}_{xy,uv}$ approach zero, and all $G^{(P)}_{yx,uv}$  approach $K^{yx,uv}_{1 \rightarrow 2}(0)$.
Thus
\begin{equation}
\begin{aligned}
\hat{H}_{1 \leftrightarrow 2}&=\frac{\mu_0}{4\pi r^3}[\hat{m}_1 \cdot \hat{m}_2-3(\hat{m}_1 \cdot \hat{e}_{\mathbf{r}})(\hat{m}_2 \cdot \hat{e}_{\mathbf{r}})],
\end{aligned}
\end{equation}
which could be written in the form $\lambda[3(\mathbf{I}_1\cdot\mathbf{r})(\mathbf{I}_2\cdot\mathbf{r})-\mathbf{I}_1\cdot\mathbf{I}_2]$, this is just the approximation $H_{dd}$ used in the main text. 
From $H_{dd}$ one further obtains $H_{zz}$. 
The initial state $1/2(\left|\uparrow\uparrow\right\rangle+\left|\uparrow\downarrow\right\rangle+\left|\downarrow\uparrow\right\rangle+\left|\downarrow\downarrow\right\rangle)$ evolves to $1/2(e^{-i\phi_{\uparrow\uparrow}}\left|\uparrow\uparrow\right\rangle+\left|\uparrow\downarrow\right\rangle+\left|\downarrow\uparrow\right\rangle+e^{-i\phi_{\downarrow \downarrow}}\left|\downarrow\downarrow\right\rangle$ under $H_{zz}$. 
This could be described as resulting from the phase accumulation in different branches by a quantum-controlled classical field, without giving a Hilbert space to the field. The accumulated phases in each branch could be calculated by classical magnetic dipole interaction. Thus we see that due to $r/c$ being small, the QFT description and the quantum-controlled classical field one is indistinguishable.  

\section*{Acknowledgement} This work is supported by the Key-Area Research and Development Program of Guang Dong Province under Grant No.2019B030330001, the National Natural Science Foundation of China under Grant No.11774436, No.11974434 and No. 12074439, Natural Science Foundation of Guangdong Province under Grant 2020A1515011159, Science and Technology Program of Guangzhou, China 202102080380, the Central -leading-local Scientific and Technological Development Foundation 2021Szvup172. Ji Bian receives support from China Postdoctoral Science Foundation under Grant No.2021M703768. Le Luo acknowledges the support from Guangdong Province Youth Talent Program under Grant No.2017GC010656. The authors are thankful to Dr. Rongxin Miao and Dr. Longlong Feng for their discussion. \\


\end{document}